\documentclass[prd,eqsecnum,twocolumn,amsfonts,amssymb]{revtex4}

\usepackage{graphicx}

\usepackage{amsmath}

\usepackage{bm}

\newcommand{\beq}{\begin{equation}}
\newcommand{\eeq}{\end{equation}}
\newcommand{\beqs}{\begin{eqnarray}}
\newcommand{\eeqs}{\end{eqnarray}}

\begin{document}

\title{Duality in a Supersymmetric Gauge Theory From a Perturbative Viewpoint}

\author{Thomas A. Ryttov$^a$ and Robert Shrock$^b$}

\affiliation{(a) \ CP$^3$-Origins, University of Southern Denmark, 
 Campusvej 55, Odense, Denmark}

\affiliation{(b) \ C. N. Yang Institute for Theoretical Physics and
Department of Physics and Astronomy, \\
Stony Brook University, Stony Brook, NY 11794, USA }

\begin{abstract}

  We study duality in $\mathcal{N}=1$ supersymmetric QCD in the non-Abelian
  Coulomb phase, order-by-order in scheme-independent series expansions. Using
  exact results, we show how the dimensions of various fundamental and
  composite chiral superfields, and the quantities $a$, $a/c$, and $b$ at
  superconformal fixed points of the renormalization group emerge in
  scheme-independent series expansions in the electric and magnetic
  theories. We further demonstrate that truncations of these series expansions
  to modest order yield very accurate approximations to these quantities.

\end{abstract}

\maketitle

% =======================================================================

\section{Introduction}
\label{intro_section}

Transformations that allow one to deal with a strongly coupled quantum field
theory as a weakly coupled field theory in a different form have proved to be
very powerful throughout the history of physics. An important example is
provided by the lattice formulation of quantum chromodynamics (QCD).  Although
the property of asymptotic freedom made possible perturbative calculations at
large Euclidean energy/momentum scales $\mu$ in the deep ultraviolet (UV), the
growth of the running gauge coupling $g(\mu)$ in the infrared (IR) prevented
reliable perturbative calculations at low energies.  However, this difficulty
was surmounted by Wilson's formulation of the theory on a (Euclidean) lattice
\cite{wilson74}, in which the plaquette term in the action is multiplied by the
coefficient $\beta = 2N_c/g_0^2$, where $g_0$ is the bare gauge coupling. Thus,
the strong coupling limit $g_0 \to \infty$ is equivalent to $\beta \to 0$,
allowing analytic strong-coupling Taylor series expansions in powers of
$\beta$.  By means of such an expansion, the area-law behavior of the Wilson
loop and hence confinement in QCD were proved for strong $g_0$
\cite{wilson74}. In a different but related way, a duality transformation links
two different regimes of a theory. In statistical mechanics, for a
two-dimensional Ising model, a duality transformation maps the high-temperature
regime to the low-temperature regime and led to the calculation of 
the critical temperature in this model \cite{kramers_wannier}. Another
realization of duality occurs in the generalization of electromagnetic
theory to include Dirac monopoles.

Here we will consider a theory for which duality relations have been very
useful, namely an asymptotically free, vectorial, gauge theory (in $d=4$
spacetime dimensions, at zero temperature) with ${\cal N}=1$ supersymmetry,
having a gauge group SU($N_c$) and $N_f$ massless chiral superfields $Q^i$ and
$\tilde Q_i$, $i=1,...,N_f$, transforming in the fundamental and conjugate
fundamental representations of SU($N_c$), respectively. This theory is
invariant under a global symmetry group $G_{gb}={\rm SU}(N_f) \otimes {\rm
  SU}(N_f) \otimes {\rm U}(1)_B \otimes {\rm U}(1)_R$ symmetry, with the
representations indicated in Table \ref{matter}. Following common terminology,
we call this theory supersymmetric quantum chromodynamics (SQCD). We denote
$\alpha=g^2/(4\pi)$.  This theory has the appeal that many of its properties
are well understood at a nonperturbative level
\cite{nsvz}-\cite{shifman_review}. The property of asymptotic freedom requires
that $N_f < 3N_c$. In this range, the theory is weakly interacting in the deep
UV,  so one can self-consistently calculate its properties perturbatively.  One
can then investigate how it evolves (``flows'') from the deep UV to the IR
limit as $\mu \to 0$.  For $N_f$ slightly less than $3N_c$, there is convincing
evidence that the theory evolves to an infrared fixed point (IRFP) of the
renormalization group (RG) at $\alpha_{IR}$, at which point it is
scale-invariant and is inferred \cite{scalecon} to be (super)conformally
invariant. If $N_f$ is in the interval $I: \ (3/2)N_c < N_f < 3N_c$, the theory
flows to an IRFP in a (deconfined, chirally symmetric) non-Abelian Coulomb
phase (NACP) \cite{seiberg}.  Henceforth, we restrict our consideration to the
NACP in this theory.

At this superconformal IRFP, it was conjectured in \cite{seiberg} that the
original theory is dual to the IR limit of another ${\cal N}=1$ supersymmetric
theory with a gauge group SU($\tilde N_c$), where $\tilde N_c = N_f-N_c$, with
matter content consisting of $N_f$ chiral superfields $q_i$ and $\tilde q^i$,
$i=1,...,N_f$, in the fundamental and conjugate fundamental representations of
SU($\tilde N_c$), respectively, together with a set of $N_f^2$ gauge-singlet
``meson'' chiral superfields $\phi^i_{\phantom{i}j}$, $1 \le i,j \le N_f$. The
original and dual theories were called ``electric'' and ``magnetic'' in
\cite{seiberg,intriligator_seiberg}. The dual theory also allows for
a unique superpotential $W=\lambda \phi q\tilde{q}$, where $\lambda$ is the
superpotential coupling. Evidence for this conjectured equivalence includes the
fact that the dual theory is also invariant under the same global $G_{gb}$
symmetry, and satisfies 't Hooft anomaly matching
\cite{seiberg}-\cite{shifman_review}.

\begin{table}
\begin{center}
\begin{tabular}{c||ccccc}
&\ SU($N_c$)&\ ${\rm SU}(N_f)$ &\ ${\rm SU}(N_f)$ &\ U(1)$_B$ &\ U(1)$_R$ \\
\hline \hline 
$ Q $ & $N_c$ & $N_f$ & $1$ & $1$ & $\frac{N_f-N_c}{N_f}$ \\
$\tilde{Q}$ & $\overline{N}_c$ & $1$ & $\overline{N}_f$ & $-1$ &
$\frac{N_f-N_c}{N_f}$
\end{tabular}
\\
\vspace{1cm}
\begin{tabular}{c||ccccc}
&\ SU($N_f-N_c$) &\ ${\rm SU}(N_f)$ &\ ${\rm SU}(N_f)$ &\ U(1)$_B$ &\ 
U(1)$_R$ \\
\hline \hline 
$ q $ & $N_c$ & $\overline{N}_f$ & $1$ & $\frac{N_c}{N_f-N_c}$ & 
$\frac{N_c}{N_f}$ \\
$\tilde{q}$ & $\overline{N}_c$ & $1$ & $N_f$ & $-\frac{N_c}{N_f-N_c}$ & 
$\frac{N_c}{N_f}$ \\
$\phi$ & 1 & $N_f$ & $\overline{N}_f$ & 0 & $\frac{2(N_f-N_c)}{N_f}$
\end{tabular}
\end{center}
\caption{The matter content of the original (electric) and dual (magnetic) 
theories.}  
\label{matter}
\end{table}

The dual theory is asymptotically free for $N_f < 3\tilde N_c$, i.e., $N_f >
(3/2)N_c$ and for $N_f$ in the interval $\tilde I: \ (3/2)\tilde N_c < N_f <
3\tilde N_c$, it flows to a superconformal IRFP in the space of gauge and
$\lambda$ couplings, at which the physics is equivalent to that in the original
theory. There is an isomorphism between the intervals $I$ and $\tilde I$ such
that the upper end of $I$ where $N_f \nearrow 3N_c$ maps to the lower end of
$\tilde I$, where $N_f \searrow (3/2)\tilde N_c$, and vice versa. The
weak-coupling region in the original theory corresponds to strong coupling in
the dual theory, and vice versa.  This duality is well supported by
nonperturbative arguments, but one gains further insight by seeing how the
duality relations emerge perturbatively.  However, a conventional perturbative
calculation, as a series expansion in powers of the gauge coupling, encounters
the difficulty that although $\alpha_{IR} \to 0$ as $N_f \nearrow 3N_c$ at the
upper end of the NACP, this theory becomes strongly coupled, and hence not
amenable to this type of perturbative approach, as $N_f \searrow (3/2)N_c$ at
the lower end of the NACP.  A similar comment applies to conventional
perturbative expansions in the magnetic theory, which has a weak gauge coupling
as $N_f \nearrow 3\tilde N_c$, but is strongly coupled as $N_f \searrow
(3/2)\tilde N_c$.  Furthermore, even in the respective regions of the electric
and magnetic theories where they are weakly coupled, perturbative expansions in
powers of gauge couplings are scheme-dependent.

Here we surmount this difficulty and present, for the first time, a
scheme-independent perturbative understanding of the duality in the non-Abelian
Coulomb phase of SQCD. An important property that we utilize is the fact that
$\alpha_{IR} \to 0$ as $N_f \nearrow 3N_c$, so that, as observed in \cite{bz}
(see also \cite{gkgg}), one can alternatively express physical quantities at an
IRFP in the NACP as series in powers of the manifestly scheme-independent
quantity $\Delta_f = 3N_c-N_f$.  As $N_f$ decreases from $3N_c$ to $(3/2)N_c$,
$\Delta_f$ increases from 0 to its maximal value of $(3/2)N_c$ in the NACP.  We
have calculated scheme-independent expansions of physical quantities such as
anomalous dimensions at an IRFP in various theories \cite{dex,dexss}. By
comparisons of finite truncations of such expansions with exactly known
expressions for operator dimensions in supersymmetric gauge theories, we have
shown that these series truncated to a modest order, such as $O(\Delta_f^4)$,
can provide quite accurate approximations to these anomalous dimensions
throughout the entire NACP \cite{dexss}. The expansion parameter in the dual
theory is $\tilde \Delta_f = 3\tilde N_c - N_f = 2N_f - 3N_c$.  We will study
dimensions of various (gauge-invariant) chiral superfield operators and of
certain quantities $a$, $a/c$, and $b$ characterizing RG flows, in both the
original and dual theories, as expansions in both $\Delta_f$ and
$\tilde\Delta_f$, and will show how various relations emerge order-by-order in
these expansions. From this analysis, we will show that a combination of
finite-order expansions in these two dual expansion parameters $\Delta_f$ and
$\tilde\Delta_f$ can yield quite accurate approximations to physical quantities
throughout the entire NACP.  Our work thus demonstrates how perturbative
calculations in the well-chosen scheme-independent expansion parameters
$\Delta_f$ and $\tilde\Delta_f$ can provide insight into results based on
abstract nonperturbative methods.

For a (gauge-invariant) quantity ${\cal P}$ at the superconformal IRFP, we 
write the expansions of ${\cal P}$ in powers of $\Delta_f$ and 
$\tilde \Delta_f$ as 
\beq
{\cal P} = \sum_{n=0}^\infty p_n \Delta_f^n 
         = \sum_{n=0}^\infty \tilde p_n \tilde\Delta_f^n \ . 
\label{pseries}
\eeq
In the first term of Eq. (\ref{pseries}), one takes $N_c$ as fixed and computes
${\cal P}$ as a function of the variable $\Delta_f$, or equivalently, $N_f$. In
the second term, for the dual theory, one takes $\tilde N_c = N_f -N_c$ as
fixed; while varying $N_f$, one can keep $\tilde N_c$ fixed by formally varying
$N_c$ oppositely to $N_f$. As noted, these expansions in (\ref{pseries}) have
the advantage, relative to conventional perturbative expansions in powers of
couplings, of being scheme-independent at each order. Furthermore, the
$n$'th-order coefficients have a definite relation to terms in conventional
perturbative expansions.  One important example is where ${\cal P}$ is the
(full) scaling dimension $D_{\cal O}$ of a physical operator ${\cal O}$, which,
in general, differs from its free-field value, due to interactions.  We write
$D_{\cal O} = \sum_{n=0}^\infty D_{{\cal O},n} \Delta_f^n$, where 
$D_{{\cal O},0}=D_{{\cal O},free}$ is the free-field dimension of ${\cal O}$. For $n
  \ge 1$, the coefficient $D_{{\cal O},n}$ depends on the terms in the beta
  function of the theory up to loop order $\ell=n+1$ and on the terms in the
  conventional expansion of $D_{\cal O}$ in powers of $\alpha_{IR}$ up to loop
  order $\ell=n$, inclusive, but does not receive contributions from
  higher-loop terms. This is a very powerful observation which can be used in
  two different ways. First, it allows for a calculation of any physical
  quantity in a manifestly scheme-independent way, order-by-order, even if one
  only knows the gauge coupling beta function and $D_{\cal O}$ to some finite
  loop order. Second, if, on the other hand, one knows an exact expression for
  a quantity ${\cal P}$ at the IRFP, then one can calculate its expansions in
  powers of $\Delta_f$ and $\tilde \Delta_f$ in Eq. (\ref{pseries}) exactly, to
  all orders, without doing any explicit loop computations.

In the original (electric) theory, gauge-singlet composite chiral
superfields include the meson-type operators 
$M^i_{\phantom{i}j} = Q^i \tilde{Q}_{j}$, and baryon- and antibaryon-type
operators 
$B^{i_1\cdots i_{N_c} } = 
\epsilon_{i_1 \cdots i_{N_c}} Q^{i_1}\cdots Q^{i_{N_c}}$ and 
$\tilde B_{i_1\cdots i_{N_c} } = \epsilon^{i_1 \cdots i_{N_c}} 
\tilde{Q}_{i_1}\cdots \tilde{Q}_{i_{N_c}}$, respectively. Similarly, in the 
dual (magnetic) theory, in addition to $\phi^i_j$, 
one has the dual baryon and antibaryon operators 
$b_{i_1 \cdots i_{\tilde N_c}} = \epsilon^{i_1 \cdots i_{\tilde N_c}}
q_{i_1} \cdots q_{i_{\tilde N_c}}$ and 
$\tilde{b}^{i_1 \cdots i_{\tilde N_c}} = 
\epsilon_{i_1 \cdots i_{\tilde N_c}} 
\tilde{q}^{i_1} \cdots \tilde{q}^{i_{\tilde N_c}}$. 
Duality dictates that the meson operators are matched in the electric and
magnetic theories, and similarly for the baryon operators and the antibaryon
operators.  Hence, the meson operators in the electric and magnetic theories
must have the same dimensions, and similarly for the baryon and antibaryon
operators. 

In a superconformal theory, the dimension $D_{\cal O}$ of a gauge-invariant
chiral superfield operator ${\cal O}$ is related to the $R$-charge, 
$R_{\cal O}$, of the operator via $D_{\cal O} = (3/2)R_{\cal O}$. 
This implies, in particular, that the scaling dimension of the
(composite) electric and (fundamental) magnetic mesons is 
\beq
D_M = D_\phi \equiv D = \frac{3(N_f-N_c)}{N_f} = \frac{3\tilde N_c}{N_f} \ .
\label{dmeson}
\eeq
Note that $D_{M,free}=2$ while $D_{\phi,free}=1$. We want to understand how
Eq. (\ref{dmeson}) emerges order-by-order in the scheme-independent expansions
in the original and dual theories.  Calculating the 
series expansion of $D$ in powers of $\Delta_f$, we find
\beq
D = 2 - \sum_{n=1}^{\infty} \left( \frac{\Delta_f}{3N_c} \right)^n \ . 
\label{dMseries}
\eeq
Equivalently, in the dual theory, we find 
\beq
D = 1 + \sum_{n=1}^{\infty}\left( \frac{\tilde \Delta_f}{3\tilde{N}_c} 
\right)^n \ . 
\label{dMdualseries}
\eeq
These series have respective radii of convergence 
$|\Delta_f| = 3N_c$ and $|\tilde\Delta_f|=3\tilde N_c$, and hence converge 
throughout the entire NACP, since the maximal values of $\Delta_f$ and 
$\tilde\Delta_f$ in the NACP are $(3/2)N_c$ and $(3/2)\tilde N_c$,
respectively \cite{nt}. The truncations of the series (\ref{dMseries}) and 
(\ref{dMdualseries}) to order $n=s$ are denoted $D_s$.

\begin{figure}
  \begin{center}
    \includegraphics[width=0.35\textwidth]{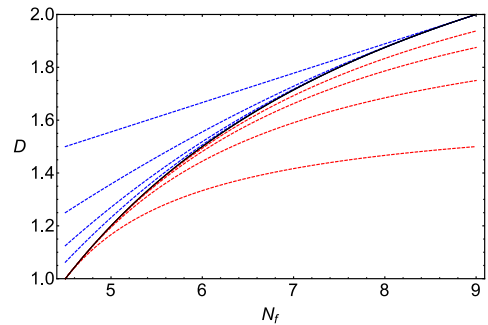}
  \end{center}
\caption{The scaling dimension $D$ of the meson chiral superfield, as 
computed to $O(\Delta_f^s)$ in the original (electric) theory (blue dashed
curves) and to $O(\tilde\Delta_f^s)$ in the dual (magnetic) theory (red
dashed curves), together with the exact $D$ (black solid curve), for the 
illustrative case $N_c=3$. From top to bottom for the blue curves and from
bottom to top for the red curves, these depict $D_s$ with $s=1,...,4$.}
\label{meson_figure}
\end{figure}

In Fig. \ref{meson_figure} we plot the values of $D$ calculated from
(\ref{dMseries}) to $O(\Delta_f^s)$, i.e., $D_s$, in the
electric theory and $D_s$ calculated from (\ref{dMdualseries})
in the magnetic theory for $1 \le s \le 4$, 
together with the exact result, for the illustrative case, $N_c=3$. Because the
coefficients of $\Delta_f^n$ and $\tilde\Delta_f^n$ in the expansions
(\ref{dMseries}) and (\ref{dMdualseries}) are positive, several monotonicity
properties follow (for $N_f$ in the NACP): (i) for fixed $s$, $D_s$ is a
monotonically increasing function of $N_f$; (ii) for fixed $N_f$, $D_s$, as
calculated in the electric theory, decreases monotonically with $s$; and (iii)
$D_s$, as calculated in the magnetic theory, increases monotonically with
$s$. As is evident from Fig. \ref{meson_figure}, the respective fractional
accuracies of the $\Delta_f$ and $\tilde\Delta_f$ series expansions in
(\ref{dMseries}) and (\ref{dMdualseries}) are highest near the upper and lower
ends of the NACP, respectively. Thus, by combining these two perturbative
calculations, we achieve an excellent approximation to the exact expression
(\ref{dmeson}) throughout all of the NACP, even with a modest value of the
truncation order, $s$ such as $s=4$. This makes use of the full power of the
duality, since it allows one to treat the strong-coupling regime in the
original theory via a perturbative calculation in the weak-coupling regime of
the dual theory, and vice versa.

We next consider the baryon-type operators. In both the electric and
magnetic theories, their scaling dimensions have to agree, and are
\beq
D_B = D_b \equiv D' = \frac{3N_c(N_f-N_c)}{2N_f} = 
\frac{3N_c \tilde N_c}{2N_f} \ .
\label{dB}
\eeq
As with the mesons, we want to understand how this expression for $D'$ emerges
order-by-order in perturbation theory, as calculated in both the original and
dual theories. In the original (electric) theory, we find
\beq
D' = N_c - \frac{N_c}{2}\sum_{n=1}^\infty 
\left ( \frac{\Delta_f}{3N_c} \right )^n \ , 
\label{dBseries}
\eeq
while in the dual (magnetic) theory we find
\beq
D' = \tilde N_c - \frac{\tilde N_c}{2}\sum_{n=1}^\infty 
\left ( \frac{\tilde\Delta_f}{3\tilde N_c} \right )^n \ . 
\label{dBdualseries}
\eeq
As before, these series converge throughout the entire NACP. 
The truncations of these series to order $n=s$ are denoted $D'_s$.  

\begin{figure}
  \begin{center}
    \includegraphics[width=0.35\textwidth]{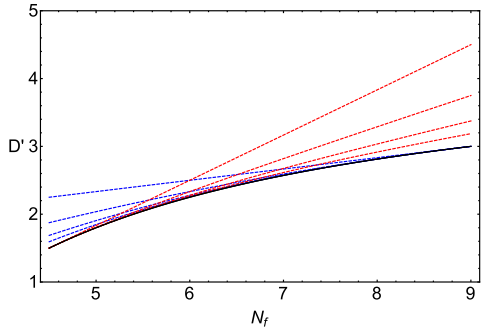}
  \end{center}
  \caption{The scaling dimension $D'$ of the baryon operators, as computed in
    the electric theory (blue dashed curves) and magnetic theory (red dashed
    curves), together with the exact result (black solid curve), for the
    illustrative value, $N_c=3$. From top to bottom for both the blue and red
    curves, these depict $D'_s$ with $s=1,...,4$.}
\label{baryon_figure}
\end{figure}

In Fig. \ref{baryon_figure} we plot the values of $D'$ calculated from
(\ref{dBseries}) to $O(\Delta_f^s)$ in the
electric theory and from (\ref{dBdualseries}) to $O(\tilde\Delta_f^s)$ in the 
magnetic theory, for $1 \le s \le 4$, 
together with the exact result. Because the coefficients of 
$\Delta_f^n$ and $\tilde\Delta_f^n$ in the
expansions (\ref{dBseries}) and (\ref{dBdualseries}) are positive, several
monotonicity properties follow for the calculations in both the electric and
magnetic theories (for $N_f$ in the NACP): (i) for fixed $s$, $D'_s$ is a
monotonically increasing function of $N_f$; and (ii) for fixed $N_f$, $D'_s$ is
a monotonically decreasing function of $s$. The same comments about respective
fractional accuracies of the $\Delta_f$ and $\tilde\Delta_f$ calculations made
for Fig. \ref{meson_figure} hold here, so that again, by combining these
two perturbative calculations, we obtain an excellent approximation to the
exact expression (\ref{dB}) throughout all of the NACP even with a modest
truncation order, $s$, such as $s=4$.

Corresponding to the global symmetry groups in $G_{gb}$, there are conserved
currents.  We will focus on the conserved current $J^{(B)}_\mu$ associated with
the U(1)$_B$ baryon number symmetry.  The two-point 
correlation function for this current (in flat space) is
\beq
\langle J^{(B)}_{\mu}(x)J^{(B)}_{\nu}(0) \rangle = 
\frac{b}{(4\pi)^2}(g_{\mu\nu}\partial^2
- \partial_\mu \partial_\nu)\frac{1}{x^4} \ . 
\label{jbjb}
\eeq
Here $b$ is a function of the couplings of the theory and changes from $b_{UV}$
to $b_{IR}$ along the RG flow. For SQCD, $b_{UV}$ is given by its respective
free-field values, $b_{UV}=2N_f/N_c$ and $b_{UV}=2N_f/\tilde N_c$ in the
electric and magnetic theories, while $b_{IR}=6$ at the IRFP in the non-Abelian
Coulomb phase in both of these theories \cite{anselmi_et_al}. Hence,
calculating $b_{UV}-b_{IR}$, we have, in the electric theory,
\beq
b_{UV}-b_{IR} = -\frac{2\Delta_f}{N_c}  \ , 
\label{bdif}
\eeq
and, in the magnetic theory, 
\beq
b_{UV}-b_{IR} = -\frac{2\tilde\Delta_f}{\tilde N_c} \ . 
\label{bdif_dual}
\eeq
These results show that higher-order contributions to $b_{IR}$ in powers of
$\Delta_f$, and equivalently in powers of $\tilde\Delta_f$, vanish. This again
shows the value of the scheme-independent series expansion method, since the
zero coefficients of the respective $\Delta_f^n$ and $\tilde\Delta_f^n$ terms
with $n \ge 2$ in Eqs.  (\ref{bdif}) and (\ref{bdif_dual}) involve complicated
cancellations when computed via conventional (scheme-dependent) series
expansions in powers of couplings.

The trace of the energy-momentum tensor in four spacetime dimensions, 
in the presence of a curved background metric $g_{\mu\nu}$, is 
\cite{traceanomaly}
\beq
T^{\mu}_{\phantom{\mu}\mu} = 
\frac{1}{(4\pi)^2} \left ( c W_{\mu\nu\rho\sigma}W^{\mu\nu\rho\sigma} - 
a E_4 \right ) 
\label{tracerel}
\eeq
where $W_{\mu\nu\rho\sigma}$ is the Weyl tensor (so
$W_{\mu\nu\rho\sigma}W^{\mu\nu\rho\sigma}=R_{\mu\nu\rho\sigma}R^{\mu\nu\rho\sigma}-2R_{\mu\nu}R^{\mu\nu}+(1/3)R^2$)
and $E_4 = R_{\mu\nu\rho\sigma}R^{\mu\nu\rho\sigma}-4R_{\mu\nu}R^{\mu\nu}+R^2$
is the Euler density, satisfying $\int d^4x \sqrt{|{\rm det}(g_{\mu\nu})|} \,
E_4 = \chi_E$, the Euler-Poincar\'e characteristic. Here,
$R_{\mu\nu\rho\sigma}$ and $R_{\mu\nu}$ are the Riemann and Ricci tensors, and
$R$ is the scalar curvature of the manifold defined by the background metric
$g_{\mu\nu}$. In $d=2$, it was proved that $c$ decreases monotonically along an
RG flow \cite{zam}, but this monotonicity does not hold in $d=4$.  The quantity
$a$ encodes important information about the flow of a quantum field theory
between two RG fixed points and satisfies the inequality that $a_{UV}-a_{IR}
> 0$ (called the $a$ theorem)
\cite{cardy}-\cite{ks},\cite{anselmi_et_al}. This is in accord with the
Wilsonian notion of thinning of degrees of freedom along an RG flow
\cite{ac,acs}. For an asymptotically free theory, $a_{UV}$ is given by the
(massless) free-field content of the theory \cite{traceanomaly}: $a_{UV} =
(1/48)(9 N_v+N_{\chi})$, where here $N_v$ and $N_\chi$ denote the numbers of
vector and chiral superfields, respectively \cite{traceanomaly,agen}. In the
original (electric) theory, $N_v = N_c^2-1$ and $N_{\chi} = 2N_cN_f$, while in
the dual (magnetic) theory, $N_v= \tilde N_c^2-1$ and $N_{\chi} = 2 \tilde
N_cN_f + N_f^2$. The duality at the superconformal IRFP dictates that the value
of $a_{IR}$ must be identical in the electric and magnetic theories; it is
\cite{anselmi_et_al} $a_{IR} = (3/16)[2N_c^2 -1 - 3(N_c^4/N_f^2)]$. 
Calculating a series expansion for $a_{IR}$ in the electric
theory, in powers of $\Delta_f$, we obtain
\beq
a_{IR} = a_{UV} - \frac{1}{144} \sum_{n=2}^\infty 
\frac{(n+1)\Delta_f^n}{(3N_c)^{n-2}} \ . 
\label{air_series}
\eeq
Eq. (\ref{air_series}) shows how the $a$ theorem is 
satisfied at each order in powers of $\Delta_f$.  In 
the magnetic theory, we find
\beq
a_{IR} = a_{UV} - \frac{1}{144}\left [ 21 \tilde \Delta_f^2 + 
\sum_{n=3}^{\infty} \frac{(n-11)\tilde\Delta_f^n}{(3 \tilde N_c)^{n-2}}
\right ] \ . 
\label{air_dualseries}
\eeq
Here again, the $a$ theorem is satisfied at each order in powers of
$\tilde\Delta_f^n$; in this case, the result follows because the leading-order
term in the square brackets, $21 \tilde\Delta_f^2$, dominates over higher-order
terms $\propto \tilde\Delta_f^n$ with $3 \le n \le 10$ with opposite sign.

\begin{figure}
  \begin{center}
    \includegraphics[width=0.35\textwidth]{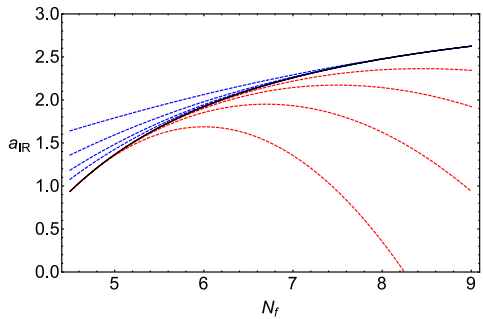}
  \end{center}
  \caption{The quantity $a_{IR}$, as computed in the electric theory (blue
    dashed curves) and magnetic theory (red dashed curves), together with the
    exact result (black solid curve), for the illustrative case $N_c=3$. From
    top to bottom, the blue curves are for $a_{IR}$ calculated to
    $O(\Delta_f^s)$ with $s=1,...,4$, and from bottom to top, the red curves
    are for $a_{IR}$ calculated to $O(\tilde\Delta_f^s)$ with $s=1,...,4$.}
\label{air_figure}
\end{figure}

In Fig. \ref{air_figure} we plot $a_{IR}$ as computed to first through fourth
order in $\Delta_f$ in the electric theory and in
$\tilde\Delta_f$ in the magnetic theory, together 
with the exact result, for the illustrative case $N_c=3$.

\begin{figure}
  \begin{center}
    \includegraphics[width=0.35\textwidth]{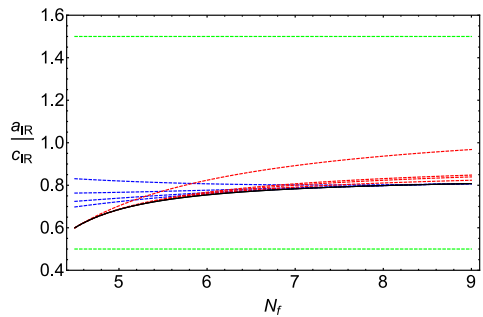}
  \end{center}
  \caption{The ratio $a_{IR}/c_{IR}$, as computed in the electric theory (blue
    dashed curves) and magnetic theory (red dashed curves), together with the
    exact result (black solid curve), for the illustrative case $N_c=3$. From
    top to bottom, the blue and red curves refer to the respective 
    calculations to $O(\Delta_f^s)$ and $O(\tilde\Delta_f^s)$ with $s=1,...,4$.
    The lower and upper green dashed lines depict the bounds 
    $1/2 \le a_{IR}/c_{IR} \le 3/2$.}
\label{acratio_figure}
\end{figure}

Although $c$ does not obey a monotonicity relation along RG flows, there is a
bound $1/2 \le a/c \le 3/2$ in a superconformal theory from the positivity of
the energy flux \cite{acratio}-\cite{acratio2017}. As with the other quantities
analyzed here, one gains insight by investigating how finite-order expansions
of this ratio at the superconformal IRFP in the original and dual theories
approach the exactly known expression.  In Fig. \ref{acratio_figure} we show
results from these $O(\Delta_f^s)$ and $O(\tilde\Delta_f^s)$ expansions
\cite{further}, together with the exact result.

In conclusion, we have shown, for the first time, how exact relations for
dimensions of chiral superfields and for the quantities $a$, $a/c$, and $b$ in
the non-Abelian Coulomb phase of SQCD emerge order-by-order in
scheme-independent perturbative series expansions, as calculated in the
original (electric) theory in powers of $\Delta_f$ and in the dual magnetic
theory in powers of $\tilde\Delta_f$.  We have demonstrated that truncated
series expansions of modest order yield quite accurate approximations to exact
results for these quantities.

\begin{acknowledgments}

The research of T.A.R. and R.S. was supported in part by the Danish National
Research Foundation grant DNRF90 to CP$^3$-Origins at SDU and by the
U.S. National Science Foundation Grant NSF-PHY-16-1620628, respectively.  

\end{acknowledgments}

% =======================================================================


\begin{thebibliography}{99}
\bibitem{nt}
%
It is understood here that $\Delta_f$ and $\tilde\Delta_f$ are always taken to
be positive, since we restrict $N_f$ to lie in the NACP.
\bibitem{ac}
%
  A thermally motivated conjecture is $f_{UV}-f_{IR} \ge 0$, with
  $f=2N_{vf}+(7/4)N_f + N_s$, where $N_{vf}$, $N_f$, and $N_s$ denote the
  number of massless vector, chiral fermion, and real scalar fields
  \cite{acs}. The inequality $f_{UV} - f_{IR} \ge 0$ is obeyed in SQCD
  \cite{acs}, and our $\Delta_f$ and $\tilde\Delta_f$ expansion methods are
  also applicable to it.
\bibitem{agen}
%
  In a general (not necessarily supersymmetric) field theory with massless
  field content consisting of $N_{vf}$ vectors, $N_f$ chiral fermions, and
  $N_s$ real scalars, $a=(1/360)[62N_{vf}+(11/2)N_f+N_s]$ \cite{traceanomaly}.
\end{thebibliography}
\end{document}